\documentclass[%
 aip,
 amsmath,amssymb,
 reprint,%
]{revtex4-2}

\usepackage{graphicx}% Include figure files
\usepackage{dcolumn}% Align table columns on decimal point
\usepackage{bm}% bold math
\usepackage[utf8]{inputenc}
\usepackage[T1]{fontenc}
\usepackage{mathptmx}
\usepackage{etoolbox}
\usepackage{float}
\usepackage{siunitx}

\usepackage[normalem]{ulem}
\usepackage{xcolor}

\newcommand{\ADDTXT}[1]{#1}
\newcommand{\REMOVETXT}[1]{\sout{}}

\makeatletter
\def\@email#1#2{%
 \endgroup
 \patchcmd{\titleblock@produce}
  {\frontmatter@RRAPformat}
  {\frontmatter@RRAPformat{\produce@RRAP{*#1\href{mailto:#2}{#2}}}\frontmatter@RRAPformat}
  {}{}
}%
\makeatother
\begin{document}

\preprint{AIP/123-QED}

\title{Equivalence of flexible stripline and coaxial cables for superconducting qubit control and readout pulses}

\author{V.~Y.~Monarkha}
\affiliation{Bluefors Oy, Arinatie 10, 00370 Helsinki, Finland}
\email{volodymyr.monarkha@gmail.com}

\author{S.~Simbierowicz}
\affiliation{Bluefors Oy, Arinatie 10, 00370 Helsinki, Finland}

\author{M.~Borrelli}
\affiliation{Bluefors Oy, Arinatie 10, 00370 Helsinki, Finland}

\author{R. ~van Gulik}%
\affiliation{Delft Circuits, Delft, The Netherlands}

\author{N. ~Drobotun}%
\affiliation{Delft Circuits, Delft, The Netherlands}

\author{D. ~Kuitenbrouwer}%
\affiliation{Delft Circuits, Delft, The Netherlands}

\author{D. ~Bouman}%
\affiliation{Delft Circuits, Delft, The Netherlands}

\author{D. ~Datta}%
\affiliation{QTF Centre of Excellence, VTT Technical Research Centre of Finland Ltd, P. O. Box 1000, 02044 VTT, Finland}

\author{P. ~Eskelinen}%
\affiliation{QTF Centre of Excellence, VTT Technical Research Centre of Finland Ltd, P. O. Box 1000, 02044 VTT, Finland}

\author{E. ~Mannila}%
\affiliation{QTF Centre of Excellence, VTT Technical Research Centre of Finland Ltd, P. O. Box 1000, 02044 VTT, Finland}

\author{J. ~Kaikkonen}%
\affiliation{QTF Centre of Excellence, VTT Technical Research Centre of Finland Ltd, P. O. Box 1000, 02044 VTT, Finland}

\author{V. ~Vesterinen}%
\affiliation{QTF Centre of Excellence, VTT Technical Research Centre of Finland Ltd, P. O. Box 1000, 02044 VTT, Finland}

\author{J. ~Govenius}%
\affiliation{QTF Centre of Excellence, VTT Technical Research Centre of Finland Ltd, P. O. Box 1000, 02044 VTT, Finland}

\author{R.~E.~Lake}
 
\affiliation{Bluefors Oy, Arinatie 10, 00370 Helsinki, Finland}

\date{\today}% It is always \today, today,
             %  but any date may be explicitly specified

\begin{abstract}
We report a comparative study on microwave control lines for a transmon qubit using: (i) flexible stripline transmission lines, and (ii) semi-rigid coaxial cables. During each experiment we performed repeated measurements of the energy relaxation and coherence times of a transmon qubit using one of the wiring configurations. Each measurement run spanned \SIrange{70}{250}{\hour} of measurement time, and four separate cooldowns were performed so that each configuration could be tested twice. From these datasets we observe that changing the microwave control lines from coaxial cables to flexible stripline transmission lines does not have a measurable effect on coherence compared to thermal cycling the system, or random coherence fluctuations.  Our results open up the possibility of large scale integration of qubit control lines with integrated component with planar layouts on flexible substrate.
\end{abstract}
%~~~~~~~~~~~~~~~~~~~~~~~~~~~~~~~~~~~~~~~~~~~~~~~~~~~~INTRODUCTION~~~~~~~~~~~~~~~~~~~~~~~~~~~~~~~~~~~~~~~~~~~~~~~~~~~~~~~~~~~~~~~~~~~~~~~~~
\maketitle
%\section*{Introduction}

Since the initial demonstration of rf single electron transistors at millikelvin temperatures \cite{schoelkopf-science-1998,Pohlen-apl-1999}, microwave-frequency control and readout has become ubiquitous in the field of engineered quantum systems and solid-state quantum bits \cite{bardin-IEEE-2021}. Signal delivery and interconnects for basic research have relied on commercially available components from other industries \cite{bardin-IEEE-2021}. Fortuitously, circuits with microwave transitions between discrete energy levels also turn out to be robust enough against thermal noise in the sample space of a dilution refrigerator so that many existing microwave technologies can be exploited for quantum control.

In particular, semi-rigid coaxial cables provide a means of signal delivery that is shielded from external electromagnetic fields, while careful selection of cable materials (i.e. cupronickel, stainless steel, or beryllium copper) \cite{weinreb-report-1982,Kurpiers-epj-2017,Krinner-epj-2019} avoids excessive passive heat loads compared to the cooling budget of the cryogenic system.  Therefore for the last 25 years the rapid progress in cryogenic nanoelectronics --- including quantum computing devices --- has been underpinned by infrastructure that includes semi-rigid coaxial cables that can span the temperature gradient from 300 K to 10 mK with minimal added heat loads.  The reduction in passive radiative heating further improved with the addition of microwave attenuators at specific temperature stages, and low-pass and infrared filters. 

Larger-scale quantum systems will require a multitude of physical microwave channels\ADDTXT{.}\REMOVETXT{where current estimates indicate that between 2.5 and 5 lines per physical qubit will be required to operate fault tolerant quantum computers.} Looking forward, the ability to miniaturize and multiply microwave wiring connections by way of circuit elements embedded on flexible polymer substrates holds great promise for solving major scaling and connectivity challenges in quantum computing \cite{Reilly-npjqi-2015, krantz-apr-2019, Corcoles-procIEEE-2020}. From a practical engineering perspective, signal delivery technologies with lower cost per channel are also needed to build larger experiments. Quantum computer control lines necessitate strict technical requirements for cryogenic thermalization and noise \cite{Krinner-epj-2019}. 

Developments to increase the density and number of physical channels is an active research area where technologies are currently being pursued in the field including coaxial arrays originally intended for microwave kinetic inductance detector readout \cite{walter-ieee-2018, smith-ieeeTAS-2021}, microstrips \cite{hernandez-IEEETAS-2017,Solovyo-ieeeTAS-2021}, and striplines \cite{Tuckerman-sst-2016,Deshpande-IEEE-2019}. There is a scarcity of published data on verification of these alternative methods for control of solid-state qubits \cite{Deshpande-IEEE-2019}. The central questions relate to understanding the effect of implementing discrete coaxial components into an embedded laminar structure of the flexible substrate, and the effect of changing the dielectric materials from teflon to a polyimide-based dielectric on qubit performance. Both changes may have a detrimental effect on qubit coherence due to the possibility of an elevated drive line temperature.

%-----------------------------------------------FIG------DIAGRAM------------------------------------------------------------------------------ old {Fig_1_02.02.2024.pdf}
\begin{figure*}
\includegraphics[width=18cm]{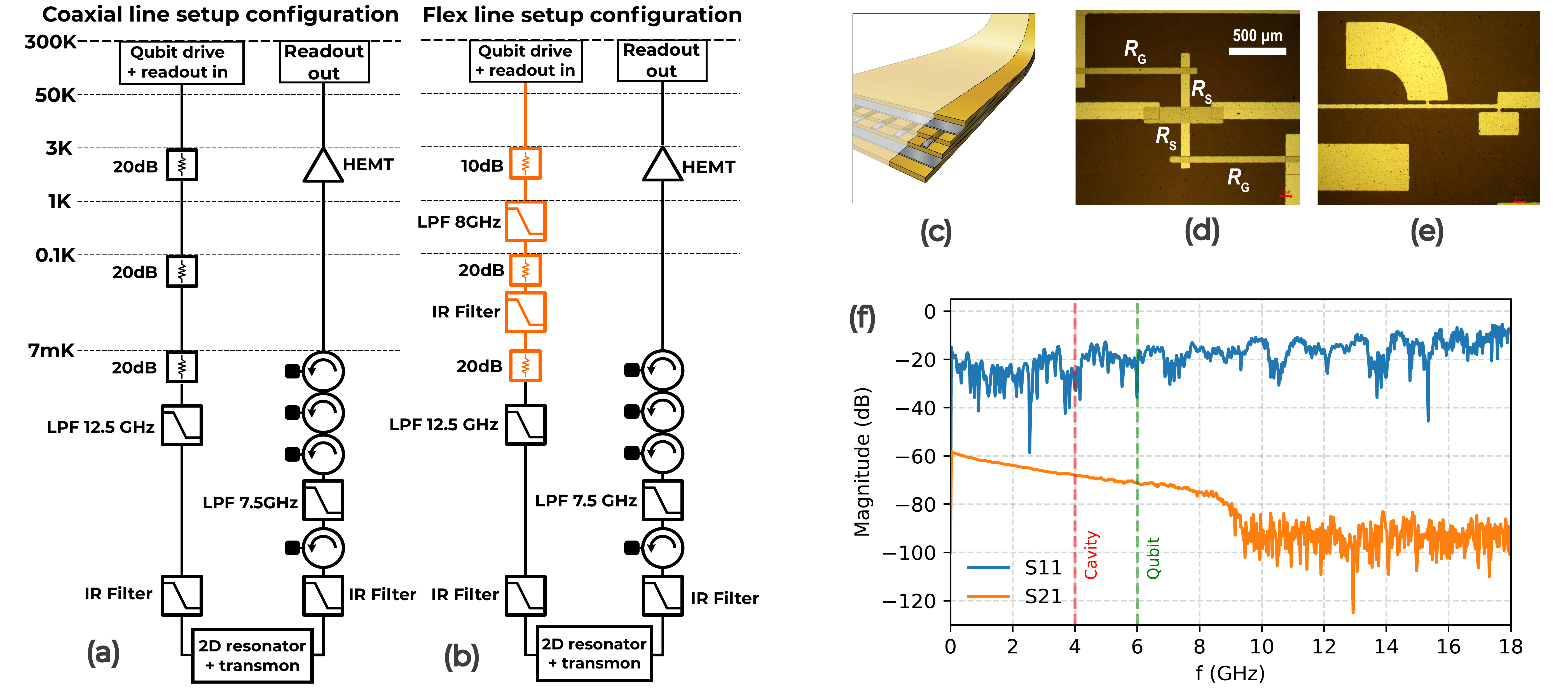}
\caption{\ADDTXT{Side-by-side comparison of the wiring configuration for the coaxial (a) and flexible stripline-based (b) input line, with the flex cable section indicated in orange. (c) Layer-by-layer schematic of the flexible stripline structure consisting of a polyimide protective outside layer, top and bottom silver groundplane, silver center conductors, and polyimide dielectric. (d) Photograph of a single T-pad 5 dB attenuator cell with indicated in-line resistances $R_S$ and resistance to ground $R_G$. (e) Photograph of a section of the stepped impedance 8 GHz low-pass filter.  (f) Representative S-parameter transmission (S21) and reflection (S11) performance of a single channel of the flex cable at a temperature gradient similar to that in (b). The S11 trace is of the room-temperature side of the flex cable.}\REMOVETXT{Comparison of setup configuration in the case of the coaxial (a) and flexible stripline-based (b) input line, transmission and reflection of the stripline-based input line measured at room temperature(c), embeded attenuator (d) and low pass filter (e) micrographs, stripline-based input line (f). Orange line indicates stripline-based path}}\label{fig:diagram} 
\end{figure*}
%--------------------------------------------------------------------------------------------------------------------------------------
%Comparison of setup configuration in the case of the coaxial (a) and flexible stripline-based (b) input line, transmission and reflection of the stripline-based input line measured at room temperature(c), embe\ADDTXT{d}ded attenuator (d) and low\ADDTXT{-}pass filter (e) micrographs, stripline-based input line (f).\ADDTXT{The }\REMOVETXT{O}\ADDTXT{o}range line \ADDTXT{in (b)} indicates \ADDTXT{the }stripline-based path
%
%-------

In this letter we provide a systematic comparison between semi-rigid cryogenic coaxial cables and flexible stripline-based cables for the application of transmon qubit drive signal delivery. To assess the two distinct signal delivery methods we compare the coherence of the qubit for each configuration in a separate cooling cycle\REMOVETXT{s} under nominally identical conditions. We measure a transmon qubit in these two configurations and observe the change in the \ADDTXT{mean}\REMOVETXT{median} of the qubit coherence times stays within the natural statistical fluctuations in coherence \cite{Burnett-npj-2019} and within changes in sample parameters due to cycling between room temperature and cryogenic environments.  In particular we wish to investigate coherence limits imposed by noise from the drive lines. In the control lines, elevated temperatures induce dephasing through resonator population and ac Stark shifts on the qubit \cite{vaaranta-pra-2022}. Our results bound coherence limitations imposed by the measurement lines and imply that stripline-based interconnects will not limit the coherence of the dispersively coupled transmon qubit we investigated.  
%\ADDTXT{as shown in Fig.~\ref{fig:diagram}(f) }
%\ADDTXT{In this letter we provide a systematic comparison between semi-rigid cryogenic coaxial cables and novel flexible stripline-based cables made by Delft Circuits for the application of transmon qubit drive signal delivery. To the best of our knowledge, this is the first time that these types of cables were compared in one experiment.}\REMOVETXT{}
%~~~~~~~~~~~~~~~~~~~~~~~~~~~~~~~~~~~~~~~~~~~~~~~~~~~~~EXPERIMENT~~~~~~~~~~~~~~~~~~~~~~~~~~~~~~~~~~~~~~~~~~~~~~~~~~~~~~~~~~~~~~~~~~~~~~~~~~~~
%\section*{Experiment}
%--- coax line discription ----
%*------ DC PART ---------
%--- flex line discription ----
The wiring configurations used in the comparative study of input wiring were installed into a Bluefors LD-250 dilution refrigerator. Each configuration is shown schematically in Fig.~\ref{fig:diagram}(a),(b).  The coaxial drive line consisted of a cryogenic SCuNi 0.86 mm diameter coaxial cables, Bluefors cryogenic attenuators with SMA interfaces, and the Bluefors IR filter as shown in Fig.~\ref{fig:diagram}(a)\cite{bluefors_coaxial_2021}. %dc part 
%The flexible stripline-based input line \ADDTXT{shown in Fig.~\ref{fig:diagram}(f) } included embedded components such as attenuators %[Fig.~\ref{fig:diagram}(d)], \ADDTXT{an }8 GHz low\ADDTXT{-}passfilter \REMOVETXT{[Fig.~\ref{fig:diagram}(e)]} and an embedded %infrared filter. \ADDTXT{Fig.~\ref{fig:diagram}(c) presents insertion loss and reflection of an flexible stripline-based input line %and the embedded components.}
%\ADDTXT{The c}\REMOVETXT{C}onductive layer of the stripline is made of \ADDTXT{a} Ag thin film with a thickness on the order of a few %microns. The thickness is a compromise between passive heat load and transmission loss. The attenuator's layout is based on a classical T-type resistive network of thin-film resistors \ADDTXT{as shown on the micrograph in Fig.~\ref{fig:diagram}(d)}. The attenuators are made with 5 dB cells to reduce the noise temperature \cite{uwave_att_2017}. The \ADDTXT{micrograph in Fig.~\ref{fig:diagram}(e) shows that the embedded} low-pass filter is based on the variation of stepped impedance layout (open circuited stubs). The IR filter is based on a lossy transmission line. The dielectric properties of the material around the IR filter structure \ADDTXT{were}\REMOVETXT{was} modified by injecting suspended metal powder to the stripline stack-up. The channel pitch of the multichannel flexible stripline is 1 mm.
%*Updated part from DC:
\ADDTXT{ Stripline is a planar type of transmission line with a thin conducting strip centered between two wide conducting ground planes with the region between the ground planes filled with a dielectric material\cite{Pozar}}\ADDTXT{The stack-up structure of the standard Cri/oFlex® Microwave Drive flexible stripline-based platform from Delft Circuits is shown in Fig. 1(c). The stripline structure consists of several micron silver film for the groundplanes and center conductor and polyimide for the dielectric, and is designed for 50 Ohm impedance. The thickness of the metal layers is a compromise between passive heat load and transmission loss. Attenuators [Fig. 1(d)], low-pass [Fig. 1(e)], and infrared filters are integrated into the planar structure of the flex to allow for a high channel density, resulting in a fully planar flexible input line. The temperature stable resistive film-based attenuator design consists of multiple 5 dB cells in order to reduce the noise temperature\cite{uwave_att_2017}. The low-pass filter is a stepped impedance type reflective filter, the large metal areas act as the capacitors and the long narrow lines in between are the inductors of the stepped impedance design. The IR filter is created by embedding suspended metal powder into the dielectric to increase the microwave losses. A standard flex has eight embedded channels with 1 mm pitch, but only one of them is used for this experiment. Thermalization of the flex cable is achieved by copper clamps at every temperature stage. 
The microwave performance of the flex cable is shown in Fig. 1(f), showing a total attenuation of 57 dB close to zero frequency. This consists mainly of the integrated attenuators totalling 50 dB, and some additional attenuation due to the resistive nature of the center conductor in the flex cable. The particular configuration of the flex cable has been chosen such that the total attenuation closely matches that of three 20 dB attenuators in the coaxial configuration.
}\REMOVETXT{The flexible stripline-based input line included embedded components such as attenuators [Fig.~\ref{fig:diagram}(d)], 8 GHz low pass filter [Fig.~\ref{fig:diagram}(e)] and an embedded infrared filter. 
Conductive layer of the stripline is made of Ag thin film with a thickness on the order of a few microns. The thickness is a compromise between passive heat load and transmission loss. The attenuator's layout is based on a classical T-type resistive network of thin-film resistors. The attenuators are made with 5 dB cells to reduce the noise temperature \cite{uwave_att_2017}. The low-pass filter is based on the variation of stepped impedance layout (open circuited stubs). The IR filter is based on a lossy transmission line. The dielectric properties of the material around the IR filter structure was modified by injecting suspended metal powder to the stripline stack-up. The channel pitch of the multichannel flexible stripline is 1 mm.}
%*------ DC PART END -----

As seen in Fig.~\ref{fig:diagram}(a),(b) for both cases (coaxial and flex) our apparatus represents a simplified superconducting qubit measurement setup that includes one input line and one output line. The setup was designed to interrogate a reference sample consisting of a fixed-frequency xmon type transmon superconducting qubit that uses an on-chip hanger style readout resonator, i.e., both control and readout pulses are delivered to the qubit through the  common input line. The fabrication technology of the qubit is niobium electrodes and ground plane on high-resistivity silicon, and multi-angle evaporated Al-AlOx-Al Josephson junctions. \ADDTXT{We estimate $E_j/E_c$ ratio to be around 60 from $E_c/h = 285 MHz$ value of an identical sample and qubit transition frequency. }

Spectroscopy experiments revealed $\omega_{\textrm{q}}/2\pi$ = 6.1 GHz, anhmarmonicity $\alpha = -260$ MHz, resonator readout \ADDTXT{frequency} $\omega_{\textrm{r}}/2\pi$ = 4.78 GHz \REMOVETXT{with a linewidth of $\kappa/2\pi$ =  19 kHz,} and resonator--qubit coupling strength of approximately $g/2\pi$ = 100 MHz. \ADDTXT{The resonator parameters such as internal quality factor $Q_i = 249\cdot~10^{3}$, coupling quality factor $Q_c = 344\cdot~10^{3}$ and linewidth $\kappa/2\pi$ = 19 kHz were measured during the initial cooldown and are assumed to be reasonably constant due to unchanged 2D geometry of the chip.} \ADDTXT{The qubit readout measurement was performed in low power dispersive regime with a measured dispersive shift $\chi$ of 1~Mhz.} The qubit was installed within a dual-layer (high permeability layer and internal superconducting layer) magnetic shield where the inner superconducting shield layer was coated with a layer of Berkeley black absorber resin \cite{Kitzman-jltp-2022} designed to absorb radiation that could induce pair breaking in the superconducting thin films of the qubit \cite{Persky-rsi-1999}. The sample holder was thermally coupled to the flange of a dilution refrigerator cryostat cooled to a system base temperature of 7 mK. \ADDTXT{The measurements would typically start a few days after the cryostat would reach the base temperature to ensure good thermalization of all microwave components.}  

To provide a performance comparison between coaxial and flexible microstrip drive-lines we performed two rounds of measurements of interleaved T$_{1}$,T$^{*}_{2}$ with each of the two different wiring configurations. In the first round the qubit sample was characterized while being connected to a coaxial-cable-based input line [see Fig.~\ref{fig:diagram}(a)] and in the second round a flexible stripline-based input line was used instead [see Fig.~\ref{fig:diagram}(b,f)].  In both rounds the spectroscopy measurements were followed by a sequence of interleaved measurements of the longitudinal and transverse relaxation times T$_{1}$ and T$^{*}_{2}$ respectively, performed over a sufficient period of time to include the effect of long-term variation of T$_{1}$ and T$^{*}_{2}$ over time \cite{Burnett-npj-2019,li-ieee-2023}. The pi-pulse duration for T$_{1}$ and Ramsey measurement was chosen to be \SI{60}{\ns}, with \ADDTXT{number of averages of each measurement}\REMOVETXT{readout averaging} between 4000 to 5000\REMOVETXT{samples}. \ADDTXT{Due to weak coupling of the readout resonator to the on-chip transmission line the averaging time to achieve sufficient SNR was as long as 16 minutes for a combined T$_{1}$, Ramsey T$^{*}_{2}$ measurement cycle.} To account for how the qubit sample parameters change due to thermal cycling, \REMOVETXT{multiple}\ADDTXT{two} warm-up/cool-down cycles were performed for each round.

%In this application we utilize this critical photon number $n_{c} = \Delta / (4g^{2})$ to estimate the difference in attenuation of the input lines at frequency $\omega_{\textrm{r}}$.  $P(n,Q)..$.  $(~nc?)$
We also measured onset of the high-transmission “bright” state to determine the critical power, which depends sensitively on the initial qubit state \cite{reed-prl-2010,blais-rmp-2021}.  The estimated difference of 3.8 dB between the flexible drive line and coaxial one is in agreement with \ADDTXT{the }value obtained from comparing calibrated pi-pulse amplitudes used in the T$_{1}$,T$^{*}_{2}$ measurement sequence.

%Show stability plot (T1, T2*, Tphi, Fqb),
%Histograms,
%Allen deviation plots. (white/pink noise fit)
%(optionally noise density, fit with same parameters)
%------------------------------------------------TABLE----QUBIT-PARAMETER-STATISTICS-------------------------------------------------
\begin{table*}
\caption{\label{tab:cooldowns}Sample parameter reproducibility over cooldown-warmup cycle in chronological order. \ADDTXT{The table shows drive line type, qubit tranition frequency, mean values of T$_{1}$ and T$^{*}_{2}$, corresponding values of standard deviation $\sigma_{~T_{1}}$ and $\sigma_{~T_{2}}$, mean pure dephasing time, number of interleaved measurement repetitions leading to the average value and estimated equivalent noise temperature.} }
\begin{ruledtabular}
\begin{tabular}{ccccccccccc}
% &\multicolumn{6}{c}{Return loss (dB)}\\
\textbf{Cooldown} &Drive~line&\ADDTXT{$\omega_{\textrm{q}}/2\pi$}\REMOVETXT{F$_{ge}$}&Mean T$_{1}$&$\sigma_{~T_{1}}$&Mean T$^{*}_{2}$&$\sigma_{~T_{2}}$&Mean T$_{\phi}$&Repetitions\REMOVETXT{Points}& Total duration&Equivalent noise T\\ \hline
\\[-0.8em]
\textbf{A$^{\dag}$} &Coax&6.0555~GHz&46 $\mu$s&13.3 $\mu$s&48 $\mu$s&13.3 $\mu$s&100 $\mu$s&373&93~hours&89 mK   \\ 
\\[-0.8em]
\textbf{B} &Coax&6.0588~GHz&46 $\mu$s&16.2 $\mu$s&49 $\mu$s&19.8 $\mu$s&105 $\mu$s&286&71~hours&87 mK     \\ 
\\[-0.8em]
\textbf{C} &Flex&6.06081~GHz&47 $\mu$s&15.7 $\mu$s&62 $\mu$s&21.6 $\mu$s&182 $\mu$s&1000&252~hours&71 mK     \\ 
\\[-0.8em]
\textbf{D$^{\dag\dag}$} &Flex&6.07349~GHz&42 $\mu$s&12.9 $\mu$s&47 $\mu$s&18.2 $\mu$s&106 $\mu$s&714&177~hours&91 mK   \\ 
\\[-0.8em]
\end{tabular}
\end{ruledtabular}
\begin{tabular}{l}
\textbf \dag ~for cooldown A the IR filter and a single junction isolator following the sample were absent.
\end{tabular}
\begin{tabular}{l}
\textbf \dag \dag ~analyzed measurement sequence contains \ADDTXT{a brief unintentional }interruption\REMOVETXT{s}.
\end{tabular}
%&\multicolumn{6}{c}{Return loss (dB)}\\
\end{table*}
%~~~~~~~~~~~~~~~~~~~~~~~~~~~~~~~~~~~~~~~~~~~~~~~~~~~~~~~~RESULTS~~~~~~~~~~~~~~~~~~~~~~~~~~~~~~~~~~~~~~~~~~~~~~~~~~~~~~~~~~~~~~~~~~

%\section*{Results}
%[Choosing the right measurement run duration]
%In order to confirm a good duration time for a sequence of 

\begin{figure}
\includegraphics[width=8.5cm]{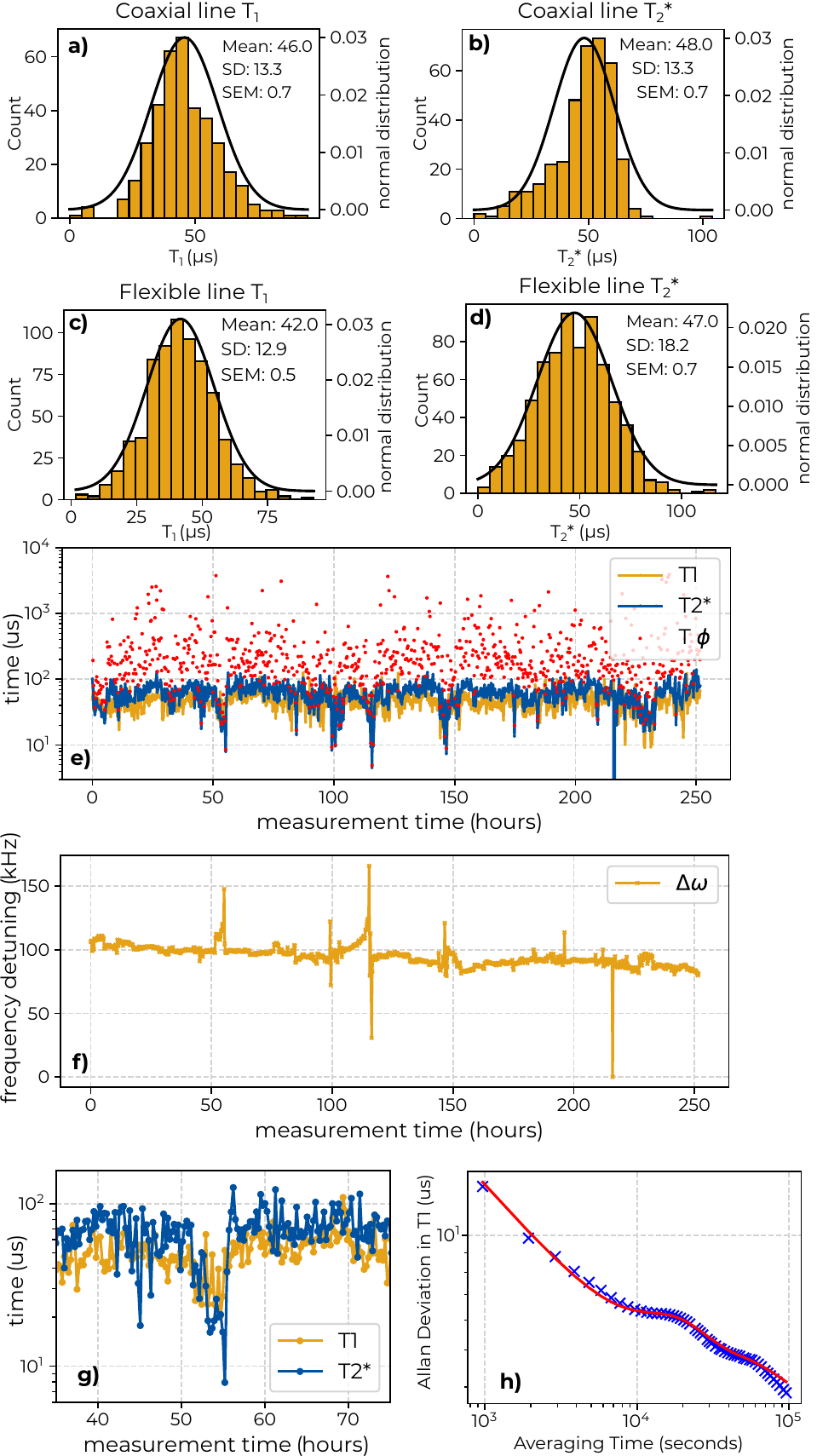}
\caption{Histograms of qubit relaxation times and coherence times for coaxial wiring (a,b) and flexible stripline wiring (c,d) for cooldowns A and D \ADDTXT{showing mean values, standard deviation and standard error in mean. Black line shows a fit of the normal distribution to the histogram}. Time evolution of T$_{1}$,T$^{*}_{2}$ and T$_{\phi}$ (e) as well as the qubit frequency stability (f) for a long interleaved measurement sequence with flexible stripline wiring as drive line. Example period of suppressed T$_{1}$ and T$^{*}_{2}$ values (g). Allan deviation diagram of T$_{1}$ fluctuations \ADDTXT{(h)}. }\label{fig:coherence_hist} 
\end{figure}

A summary of the measurement results is displayed in Fig.~\ref{fig:coherence_hist}. The interleaved T$_{1}$ and T$^{*}_{2}$ data were acquired in four long measurement runs that lasted 70--250 hours; each cooldown is indicated with a letter A--D in Table~\ref{tab:cooldowns}. Data from cooldowns A--D were acquired in chronological order several weeks and the the \ADDTXT{mean}\REMOVETXT{median} $T_1$ values were \SIlist{46;46;47;42}{\us}. In contrast, the \ADDTXT{mean}\REMOVETXT{median} T$^{*}_{2}$ values were \SIlist{48;49;62;47}{\us}. \ADDTXT{No deliberate filtering or removal of the data was done.}
 Examples of the distributions of energy relaxation time and coherence times are shown in Figs.\ref{fig:coherence_hist}(a),(b) for cooldown A, and Figs.\ref{fig:coherence_hist}(c),(d) for cooldown D. The time evolution of T$_{1}$,T$^{*}_{2}$ and $T_{\phi}$ for cooldown C can be seen in Fig.~\ref{fig:coherence_hist}(e), and time evolution data of qubit sample transition frequency extracted from the Ramsey measurements can be seen in Fig.~\ref{fig:coherence_hist}(f). \ADDTXT{The dependence shows a noticeable drift over the course of 250 hours that we speculate to be related to long term TLS evolution.}
The data in Fig.~\ref{fig:coherence_hist}(e) show periods of suppressed T$_{1}$ and T$^{*}_{2}$ that last up to approximately 5 hours motivating the relatively long measurement runs described above. Figure~\ref{fig:coherence_hist}(g) shows a zoom-in of one such period of reduced qubit lifetime and coherence. Thanks to significant number of \ADDTXT{repetitions}\REMOVETXT{points acquired}, we can analyze the time scales of fluctuations in T$^{*}_{2}$ from cooldown B [Fig.~\ref{fig:coherence_hist}(h)] using an Allan deviation diagram  following the method introduced in Ref.~\cite{Burnett-npj-2019,li-ieee-2023}. However the attempt to fit the $T_1$ Allan deviation diagram with a mixture of white, $1/f$ and Lorentzian noise (following the approach in Ref.~\citenum{Burnett-npj-2019}) did not show a possibility of a good fit with our data. The best fit result was achieved assuming the noise to be a mixture of white, $1/f$ and band limited noise (white noise band enveloped by two Heaviside functions in frequency domain).

%To account for possible effects of qubit sample parameter change due to cycling between cryogenic environment  conditions and room environment conditions we performed multiple cooldowns with unchanged wiring configuration for both coaxial and flexible drive lines. The results of several cooldowns are presented in Table~\ref{tab:cooldowns}. %and Welch method noise power spectrum

%-----------------------------------------------------FIG----HISTOGRAMS-------------------------------------------------------
%-----------------------------------------------------------------------------------------------------------------------------

%\section*{Discussion}
%~~~~~~~~~~~~~~~~~~~~~~~~~~~~~~~~~~~~~~~~~~~~~~~~~~~~~~~~Discussion~~~~~~~~~~~~~~~~~~~~~~~~~~~~~~~~~~~~~~~~~~~~~~~~~~~~~~~~~~~~~~~~~~

The goal of the study was to correlate the changes in the lifetime and coherence of the transmon qubit to the changes made to the wiring configuration. However, in order to understand the influence of the measurement lines on qubit dynamics, we first briefly discuss various contributions to loss of coherence for transmons. The fundamental theoretical limit on the qubit sample phase coherence time T$^{*}_{2}$ is the doubled transverse (or energy) relaxation time T$_{1}$. T$_{1}$ is governed by coupling to (loss to) input-output (i/o) lines and coupling to two-level systems (TLSs). The chosen sample has weak coupling to the on-chip transmission line minimizing the energy decay rate in to the line: no Purcell filter was utilized in the sample. Regarding coupling to the TLS bath we observe telegraphic noise in qubit transition frequency time evolution which is an unambiguous sign of interaction with TLS. We also see peaks associated with TLS interaction \cite{Burnett-npj-2019} on an Allan deviation diagram presented on Fig.~\ref{fig:coherence_hist}(h).   
%best fitted with simulated noise diagram containing white, $1/f$, and low\ADDTXT{-}passband limited white noise components.

Various other factors contribute to the coherence of the qubit chip including: thermal quasiparticles, non-equilibrium quasiparticles \cite{Serniak-prl-2018} arising due to stray infrared photons originating from higher temperature stages of the cryostat \cite{corcoles_apl_2011} as well as ionizing radiation coming from the experimental setup component materials or outer environment \cite{Vepsalainen-nature-2020}. Those factors are assumed to be the same for both drive line configurations.

Finally, one of the main dephasing channels limiting phase coherence time T$^{*}_{2}$ is the residual thermal photon population in the readout cavity \cite{wang-prapp-2019}. The transverse coupling between the resonator and qubit produces a shift of 2$\chi$ in the qubit frequency for each added photon stored in the resonator causing the qubit to dephase when the photon number fluctuates \cite{koch-pra-2007,clerk-pra-2007,yeh-jap-2017,Yan-natcomm-2016,vaaranta-pra-2022}. Our experimental design emphasizes influences to qubit dynamics that arise from the control wiring. As shown in Figs.~\ref{fig:diagram}(a),(b), the qubit does not have an individual XY control line such as in Ref.~\cite{barends-prl-2013}. Instead the drive line connects to the on-chip feedline, and noise from the drive line influences the qubit primarily through changes in the readout resonator population. The experiment also isolates the contribution of the wiring by: 1) utilizing the identical transmon for both wiring configurations, and 2) collecting enough statistics to resolve shifts in the distribution within temporal fluctuations that occur naturally \cite{Burnett-npj-2019,Klimov-prl-2018}.

Our initial speculation was that changing drive line construction from coaxial to flexible microstrip lines would lead to a significant difference in qubit lifetime and coherence.  For example, changing the wiring configurations varies both dielectric and conductor material, transmission line type, signal conditioning components including attenuators and filters, and drive line thermalization pathways. We expected to observe these differences in the extracted cumulants in the distributions of repeated measurements report in Table~\ref{tab:cooldowns}.

%\begin{figure}
%\includegraphics[width=8.5cm]{Fig_3_sem.pdf}
%\caption{Standard error in mean (SEM) for T$_{1}$ and T$^{*}_{2}$ depending on the duration of the enveloped %measurement sequence (data for cooldown C). \label{fig:equivalent_temperature}}
%\end{figure}
In contrast to the initial hypothesis, we observed no repeatable change in the measured qubit properties due to the changes in wiring configuration.  Although \ADDTXT{mean}\REMOVETXT{median} coherence times T$^{*}_{2}$ increase between cooldown B and C from \SI{49}{\us} to \SI{62}{\us}, the effect is not repeatable in cooldown D after the thermal cycle. 
The crucial point in this article is the observation of only a small change in the \ADDTXT{mean}\REMOVETXT{median} of the dephasing rates in Fig.~\ref{fig:coherence_hist} when changing the drive line construction entirely.  The \ADDTXT{mean}\REMOVETXT{median} coherence values fluctuate between cooldowns, but  this effect appears not to be caused by the vastly different wiring configurations in cooldowns A and B, versus those in cooldowns C and D. Our results suggest that changing to a flexible stripline-based drive line does not limit coherence of the transmon below a value of \SI{60}{\us}. We would expect contributions to dephasing \ADDTXT{to }\REMOVETXT{that }arise primarily from excess population of photons in the resonator for the dispersively coupled transmon qubit \cite{uwave_att_2017}. As seen from Table~\ref{tab:cooldowns} the mean T$_{\phi}$ is\ADDTXT{ close to $100\mu$s}\REMOVETXT{very close} in both drive line configurations. Following the\REMOVETXT{theoretical} analysis in \ADDTXT{Ref.~\cite{wang-prapp-2019}}\REMOVETXT{Ref.~\cite{vaaranta-pra-2022}} we estimate dependency of the decoherence time T$^{*}_{2}$ on the system noise temperature of the drive line. We predict that the limit in coherence due to thermal noise from the drive line is \SI{90}{\us} given our parameters. Therefore, we expect both drive lines to be below an effective temperature of around \SI{90}{\milli\kelvin} \ADDTXT{(corresponding to $8\cdot~10^{-2}$ thermal photons)} to explain these observations. \ADDTXT{In principle, noise from both input and readout lines determine the total noise due to the hanger configuration of the readout resonator. However, we calculate that four isolating stages (20~dB each) thermalized to 10~mK reduce the 3~K in-band back-action noise of the HEMT down to around 15~mK with corresponding number of thermal photons on the order of $10^{-7}$. Therefore, we do not expect amplifier back-action to significantly contribute to the decoherence implying that the upper limit on coherence times is indeed imposed by the drive lines.}

\section*{Acknowledgments}
We thank Mikael Kervinen for helpful comments on the manuscript. VTT acknowledges financial support from the EU Flagship on Quantum Technology HORIZON-CL4-2022-QUANTUM-01-SGA project 101113946 OpenSuperQPlus100, Research Council of Finland Centres of Excellence program (project nos. 352934 and 352935), and a Quantum computer co-development project funded by the Finnish government.

%\clearpage

\section*{Data availability}
The data that support the findings of this study are available from the corresponding author upon reasonable request.

\section*{References}
%\nocite{*}
%\bibliography{aipsamp}% Produces the bibliography via BibTeX.

\providecommand{\noopsort}[1]{}\providecommand{\singleletter}[1]{#1}%

\end{document}